\documentclass[aps,pre,twocolumn,superscriptaddress,floatfix]{revtex4}

\usepackage{amssymb}
\usepackage{epsfig}
\usepackage{amsmath}

\usepackage{graphicx}
\usepackage{subfigure}

\begin{document}

\title{Effects of periodically modulated coupling  on
amplitude death in nonidentical oscillators}

\author{Weiqing Liu}
\email{wqliujx@gmail.com} \affiliation{School of Science, Jiangxi
University of Science and Technology, Ganzhou 341000, China}

\author{Xiaoqi Lei}
\affiliation{School of Science, Jiangxi University of Science and
Technology, Ganzhou 341000, China}

\author{Jiangnan Chen}
\affiliation{School of Information and computer engineering,
Pingxiang University, Pingxiang, 337055, China}

\begin{abstract}
 The effects of periodically modulated coupling  on
amplitude death in two coupled nonidentical oscillators are
explored. The AD domain could be significantly influenced by tuning
the modulation amplitude and the modulation frequency of the
modulated coupling strength. There is an optimal value of modulation
amplitude for the modulated coupling with which the largest AD
domain is observed in the parameter space. The AD domain is enlarged
with the decrease of the modulation frequency for a given small
modulation amplitude, while is shrunk with decrease of the
modulation frequency for a given large modulation amplitude. The
mechanism of AD in the presence of periodic modulation in the
coupling is investigated via the local condition Lyapunov exponent
of the coupled system. The stability of AD state can be well
characterized by conditional Lyapunov exponent. The coupled system
experiencing from the oscillatory state to AD is clearly indicated
by the observation that the conditional Lyapunov exponent transits
from positive to negative. Our results are helpful to many potential
applications for the research of neuroscience and dynamical control
in engineering.
\end{abstract}

 \maketitle

\newpage
\textbf{I. Introduction}

By modeling the coupled nonlinear oscillators, a rich source of
ideas and insights into understanding the emergence of
self-organized behaviors in physics, biology, chemistry and
neuroscience has been explored \cite{sun,kur,pik} during the last
few decades. Quenching of oscillation, as one of the basic
collective dynamic behaviors, has attracted many attentions in
various fields of nonlinear science since it has potential
application on the control of chaotic oscillations and stabilization
of various unstable dynamics in the aspects of mechanical
engineering \cite{son}, synthetic genetic networks \cite{kose,ull},
and laser systems \cite{kim,pra}. Two main categories are described
according to their generation mechanisms and manifestations,
amplitude death (AD) and oscillation death (OD) \cite{kos,sax}. In
AD, oscillations are suppressed to the same original homogeneous
steady state (HSS) \cite{sax}. It is mainly applied as a control
mechanism in physical and engineering systems. In contrast, OD
occurs due to a stabilization of an newborn inhomogeneous steady
state (IHSS), where the individual units stay in different branches
of the IHSS \cite{kos}. The main implications of OD are in
biological systems, since it has been interpreted as a background
mechanism of cellular differentiation \cite{kose,got,suz} and
related neurological conditions \cite{cur,zey}. AD is manifested to
transit to OD via Turing bifurcation \cite{ane}, mean-field
diffusive coupling, dynamics coupling \cite{zou}, and time delay
coupling in experimental observations \cite{kos2,her}.

Generally, there are three types of main factors influencing the
oscillation suppression phenomena as follows, (1) the parameter
mismatches between the nodes in coupled oscillators
\cite{aro,liu2,kos1}; (2) The structure of the interaction networks;
(3) The coupling schemes between interacting nodes. In Ref.
\cite{kos}, the standard oscillatory solutions are eliminated in a
large region of the parameter mismatches by establishing the
dominance of oscillation death under strong coupling in a set of
qualitatively different models of coupled oscillators (such as
genetic, membrane, Ca metabolism, and chemical oscillators). In our
previous work \cite{liu2},  AD are general regimes in a ring of
coupled oscillators with parameter mismatches and the spatial
distribution of the parameter mismatches significantly influences
the critical coupling strength needed for amplitude death. Rich
dynamics of oscillation quenching are observed in regular networks
such as all-to-all coupled oscillators \cite{erm} and diffusively
coupled oscillators \cite{yang,rub,atay}, as well as in the complex
networks such as small-world network \cite{hou} and BA network
\cite{liu4} where the topological property of network distinctly
influences the AD dynamics. Furthermore, a rich of complex
spatiotemporal patterns of coupled oscillators in networks is
explored such as transition from amplitude chimeras states to
chimera death states \cite{ann} where the population of oscillators
splits into distinct coexisting domains of spatially coherent
amplitude of oscillation (or oscillation death) and spatially
incoherent amplitude of oscillation (or oscillation death). The
transition process to oscillation death can be continuous one
\cite{yang} or discontinuous one \cite{guan,nan,ume} named as
explosive death.

 Various kinds of coupling schemes are  available for oscillation quenching in coupled oscillators, such as dynamic coupling
\cite{kon}, conjugate coupling \cite{kar}, nonlinear coupling
\cite{pra1}, gradient coupling \cite{liu3}, mean-field diffusive
coupling \cite{tan}, amplitude dependent coupling \cite{liu},
time-delay in coupling due to a finite propagation of the signal
\cite{zou1,zou2,praa}. In all these existing studies, the
interaction or coupling among the systems  works continuously or
permanently with time. However, the continuous interaction does not
always keep in many real systems such as the biological signal
transmission between synapses and mechanical control of engineering.
The strengths of synapses in neuronal networks are modified
according to external stimuli, the links in metabolic networks are
activated only during specific tasks which lead to non-static
interactions. On-off coupling, one manifestation of discontinuous
coupling, has been verified to optimize the synchronization
stability and speed \cite{chen,bus}. Schroder et al \cite{sch}
explored a scheme for synchronizing chaotic dynamical systems by
transiently uncoupling them and revealed that systems coupled only
in a fraction of their state space may transit to synchronous state
from non-synchronous state of formally full coupling interaction.
The synchronous efficiency may be improved in the aspect of control.
Periodic coupling, another time-varying coupling scheme, are also
verified to be a available candidate to maximize the network
synchronizability by properly selecting coupling frequency and
amplitude \cite{wang}.

Compared to the focus of time-varying coupling on synchronizability
of coupled system, the AD dynamics under the effects of
discontinuous coupling are rarely explored. Recently, AD was
observed theoretically \cite{praaa} and experimentally \cite{pra2}
in two coupled oscillators by introducing the time-varying coupling.
Sun et. \cite{sun} extended the study of AD under the influence of
on-off coupling and found that AD domains are enlarged in the
parameter space with a proper switching frequency and switching rate
of coupling. However, the discontinuous form of on-off switches is
sharp and difficult to realize physically owning to a finite
response time of the switcher. Therefore, it is natural to reveal
the effects and mechanisms of a kind of more flexible discontinuous
coupling (i.e. periodic coupling) on the oscillation quenching. The
main goal in this work is to investigate the effects of periodically
modulated coupling  on the emergence of AD in the coupled
nonidentical oscillators. In particular, we show that the occurrence
of AD in nonidentical oscillators with time-varying coupling can be
well characterized by the conditional Lyapunov exponent of the
coupled system.

\section{II. Models}
\begin{figure}
\includegraphics[width=9cm]{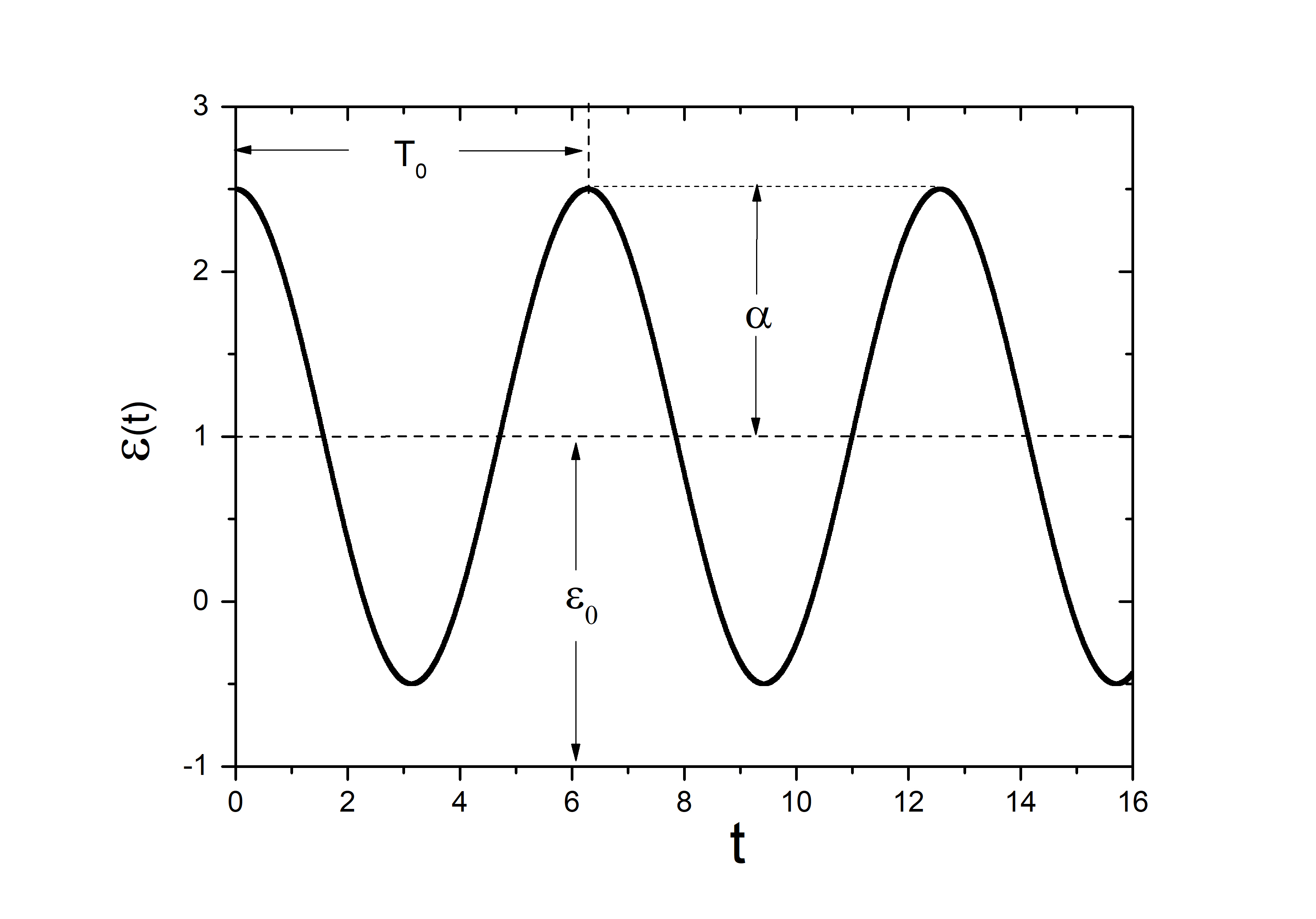}
\caption{The periodically modulated coupling strength versus time,
where $\alpha$, $\epsilon_0$, and $T_0$ are the modulation
amplitude, average coupling strength, modulation period of the
coupling term.} \label{fig_1}
\end{figure}

In this section, a periodically modulated coupling scheme is
introduced to the following system of coupled oscillators of general
form:
\begin{eqnarray}\label{eq1}
\dot{X_1}(t)&=&f_1(X_1(t))+\epsilon(t)\Gamma(X_2(t)-X_1(t)), \nonumber \\
\dot{X_2}(t)&=&f_2(X_2(t))+\epsilon(t)\Gamma(X_1(t)-X_2(t)),
\end{eqnarray}
where $X_i \in R^n (i=1,2)$, $f: R^n\rightarrow R^n$ is nonlinear
and capable of exhibiting rich dynamics such as limit cycle or
chaos, and $\Gamma$ is a constant matrix describing coupling scheme.
and $\epsilon(t)$ is a periodically modulated coupling strength as
shown in Fig. \ref{fig_1}, and can be described as,
\begin{eqnarray}\label{eq2}
\epsilon(t)=\epsilon_0 [1+\alpha cos(\omega_0 t)],
\end{eqnarray}
where $\epsilon_0$, $\omega_0$ and  $\alpha\in[0,2]$ are the average
coupling strength, modulation frequency
($\omega_0=\frac{2\pi}{T_0}$), and modulation amplitude of the
periodically modulated coupling strength, respectively. The coupling
may vary in the positive range for all time as $\alpha\in(0,1)$ and
keep constant if $\alpha=0$, otherwise the coupling strength may
vary between positive and negative, i.e. the coupling interchanges
between attractive and repulsive as $\alpha\in(1,2]$.

\section{III. Results}

\subsection{A. Coupled Stuart-Landau oscillators}
\begin{figure}
\includegraphics[width=9cm]{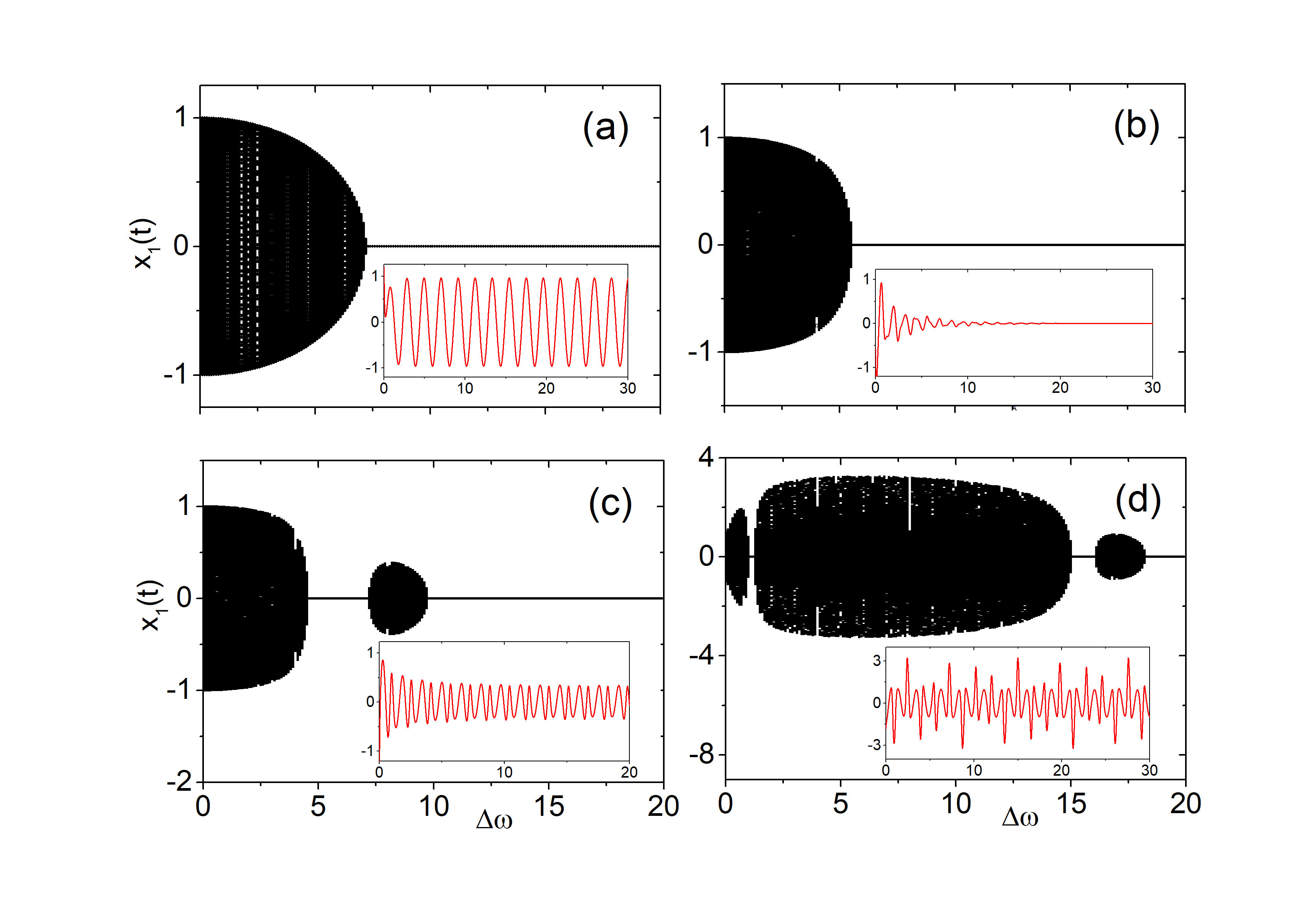}
\caption{(Color online) The bifurcation diagram of $x_1$ (gray dots)
versus frequency mismatch $\Delta\omega$ for $\epsilon_0=7$ and (a)
$\alpha=0.0$, (b) $\alpha=0.8$, (c)$\alpha=1.0$, (d) $\alpha=1.8$,
respectively. The insets are the time series of $x_1(t)$ for (a)
$\alpha=0.0$,$\Delta\omega=2$, (b)$\alpha=0.8$, $\Delta\omega=6$,
(c) $\alpha=1.0$, $\Delta\omega=8$ (d)$\alpha=1.8$,
$\Delta\omega=5$.} \label{fig_2}
\end{figure}

\begin{figure}
\includegraphics[width=9cm]{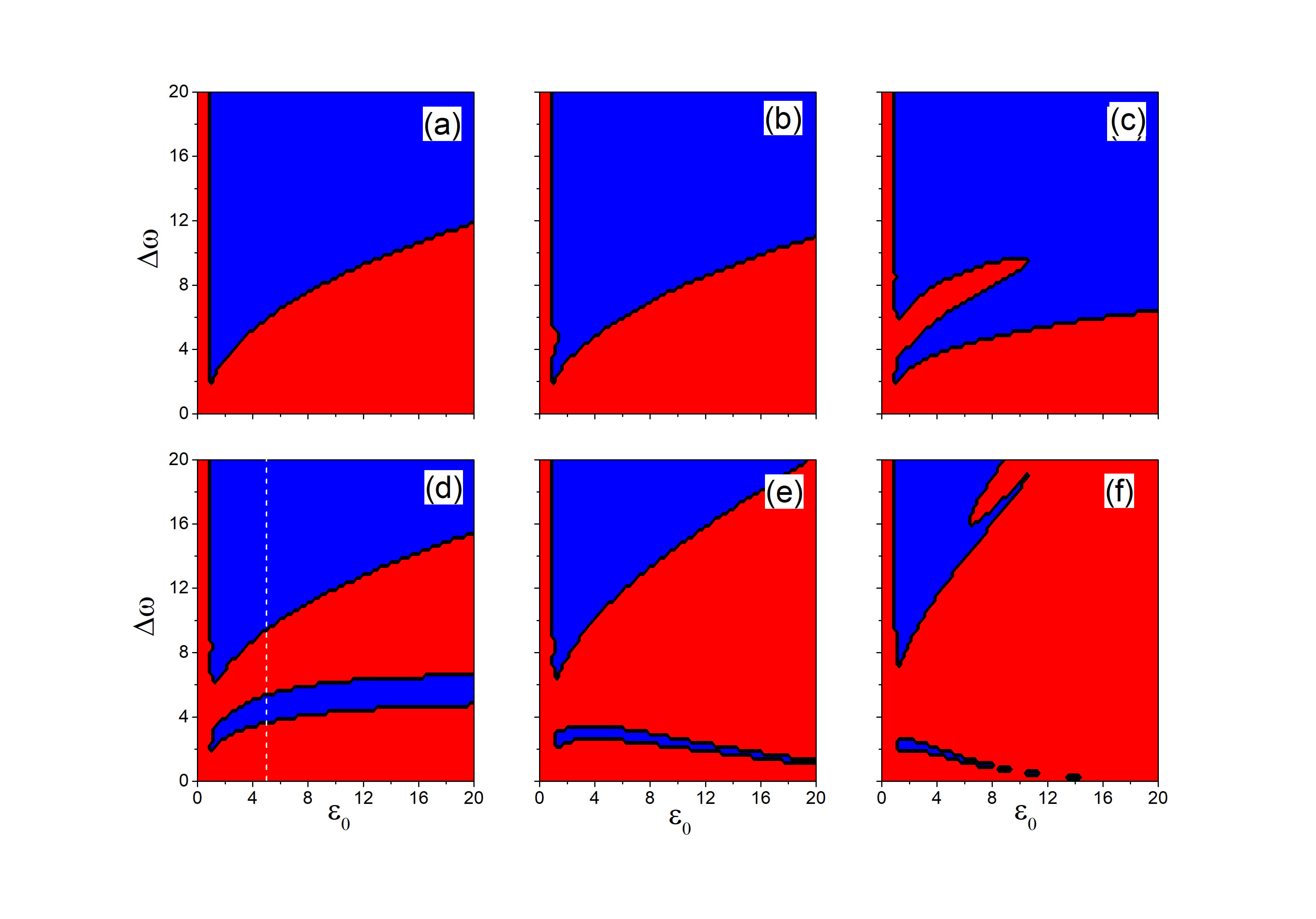}
\caption{(color online) The phase diagram of the parameters $\alpha$
versus $\Delta\omega$ for $\omega_0=4.0$ and (a) $\alpha=0.0$, (b)
$\alpha=0.5$, (c) $\alpha=1.0$, (d) $\alpha=1.1$, (e) $\alpha=1.4$,
(f) $\alpha=1.8$, respectively. Where the blue domains are in AD
states and the red domains are the oscillating states.}
\label{fig_3}
\end{figure}

In order to observe the effects of the periodically modulated
coupling on the AD dynamics, let's firstly consider the coupled
nonidentical Stuart-Landau oscillators whose dynamics can be
described as $\dot{Z_i}(t)=[1+j\omega_i-|Z_i(t)|^2]Z_i(t)$, where
$Z_i(t)=x_i(t)+jy_i(t)$, $i=1,2$,$j=\sqrt{-1}$, $\omega_i$ is the
intrinsic frequency of oscillator $i$. Without coupling ($\epsilon_0
= 0$), each oscillator has an unstable focus at the origin $|Z_i| =
0$ and an attracting limit cycle with a oscillating frequency
$\omega_{i}$. Considering the coupling scheme $\Gamma=\left[
\begin{matrix}
1&0\\
0&1
\end{matrix}\right]$ ,
AD can be stabilized in two coupled oscillators in Eq. \ref{eq1}
with frequency mismatches  $\Delta \omega=|\omega_2-\omega_1|$ for
constant coupling strength ($\alpha=0$) and $\omega_1=2$. The
results \cite{aro} indicate that the AD domain is bounded in the
interval of coupling strength for given frequency mismatches while
keeps stable when the frequency mismatch is larger than a critical
value for given coupling strength  as shown in Fig. \ref{fig_3}(a).
Since the modulated coupling strength has three control parameters,
i.e. average coupling strength $\epsilon_0$, modulation frequency
$\omega_0$, and modulation amplitude $\alpha$. Let's firstly explore
the effects of the modulation amplitude on AD with the modulation
frequency $\omega_0=4$ fixed, and the average coupling strength
$\epsilon_0=7.0$. Figs. \ref{fig_2}(a)-(d) present the bifurcation
diagram of $x_1$ for $\alpha=0.0,0.8,1.0,1.8$, respectively. For
$\alpha=0$, the coupling strength is constant, and the coupled
system transits to AD from the oscillating state (i.e. the time
series of $x_1(t)$ display periodic oscillation for
$\Delta\omega=2.0$ in Fig. \ref{fig_2}(a)) when the frequency
mismatch increases from zero to the value larger than
$\Delta\omega_{c}=7.3$. As $\alpha=0.8$, the critical value  of
$\Delta\omega_{c}$ for AD becomes $5.6$ which is less than that of
the constant coupling strength. The AD state is displayed by the
time series of $x_1(t)$ in the insets of Fig. \ref{fig_2}(b) for
$\Delta\omega=6$. The coupled system has two intervals of AD domain
$\Delta\omega\in[4.6,7.1]$ and $\Delta\omega\in[9.85,20]$ as
$\alpha=1.0$. Then, three disconnected interval of AD domain along
the direction of $\Delta\omega$ are observed as
$\Delta\omega\in[1.1,1.2]$, $\Delta\omega\in[15.1,16.0]$,
$\Delta\omega\in[18.3,20]$ for $\alpha=1.8$. The oscillating state
is in periodic 2 for $\Delta\omega=8$ and $\alpha=1.0$ and in
multi-period state for $\Delta\omega=5$ and $\alpha=1.8$ as shown in
the insets of Figs. \ref{fig_2}(c)(d),respectively.

\begin{figure}
\includegraphics[width=9cm]{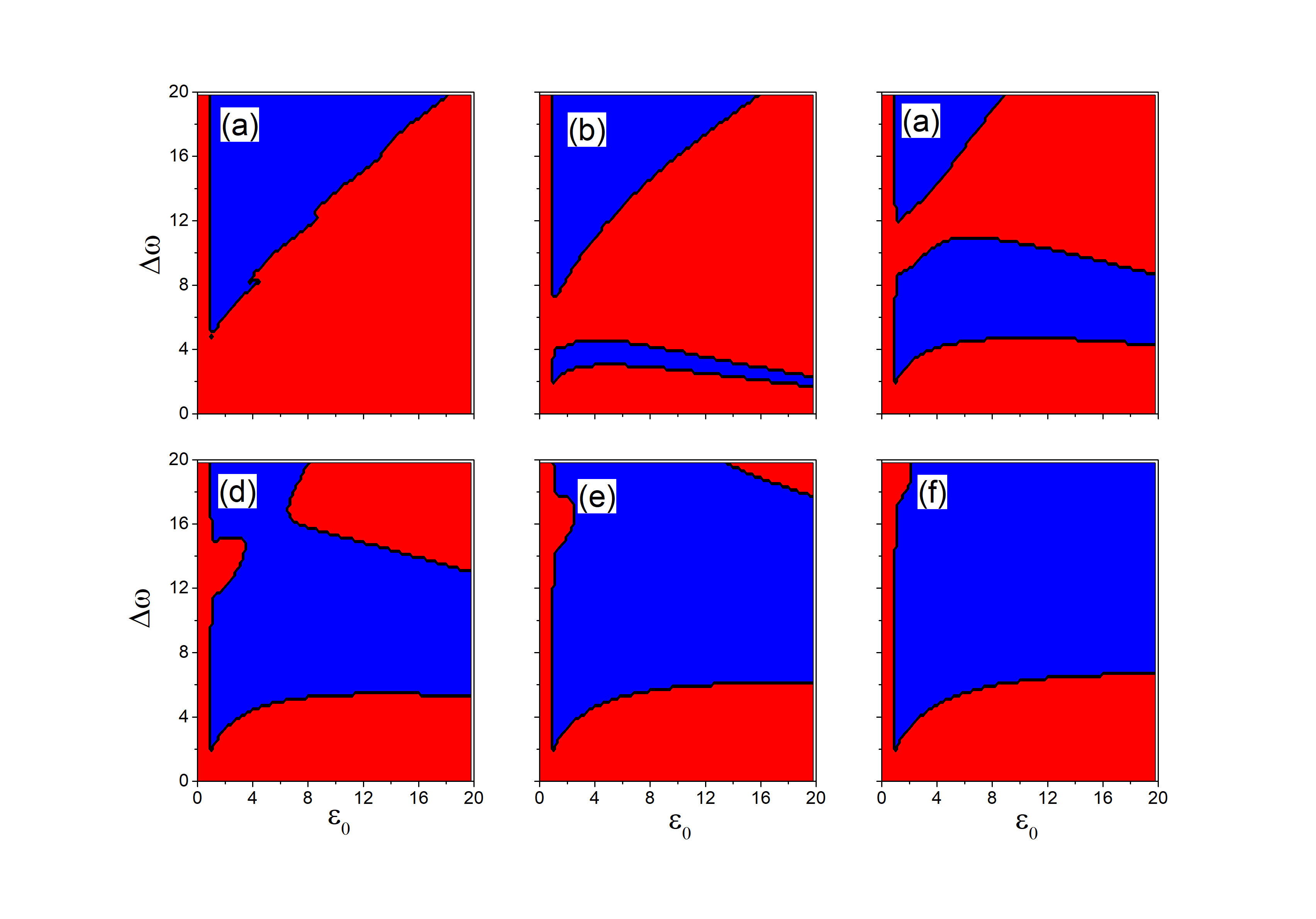}
\caption{ (color online)  The phase diagram of the parameters
$\Delta\omega$ versus $\epsilon_0$ for $\alpha=1.4$ and (a)
$\omega_0=1.0$, (b) $\omega_0=5$, (c) $\omega_0=10$, (d)
$\omega_0=13$, (e) $\omega_0=16$, (f) $\omega_0=19$, respectively.
Where the blue domains are in AD states and the red domains are the
oscillating states.} \label{fig_4}
\end{figure}

\begin{figure}
\includegraphics[width=9cm]{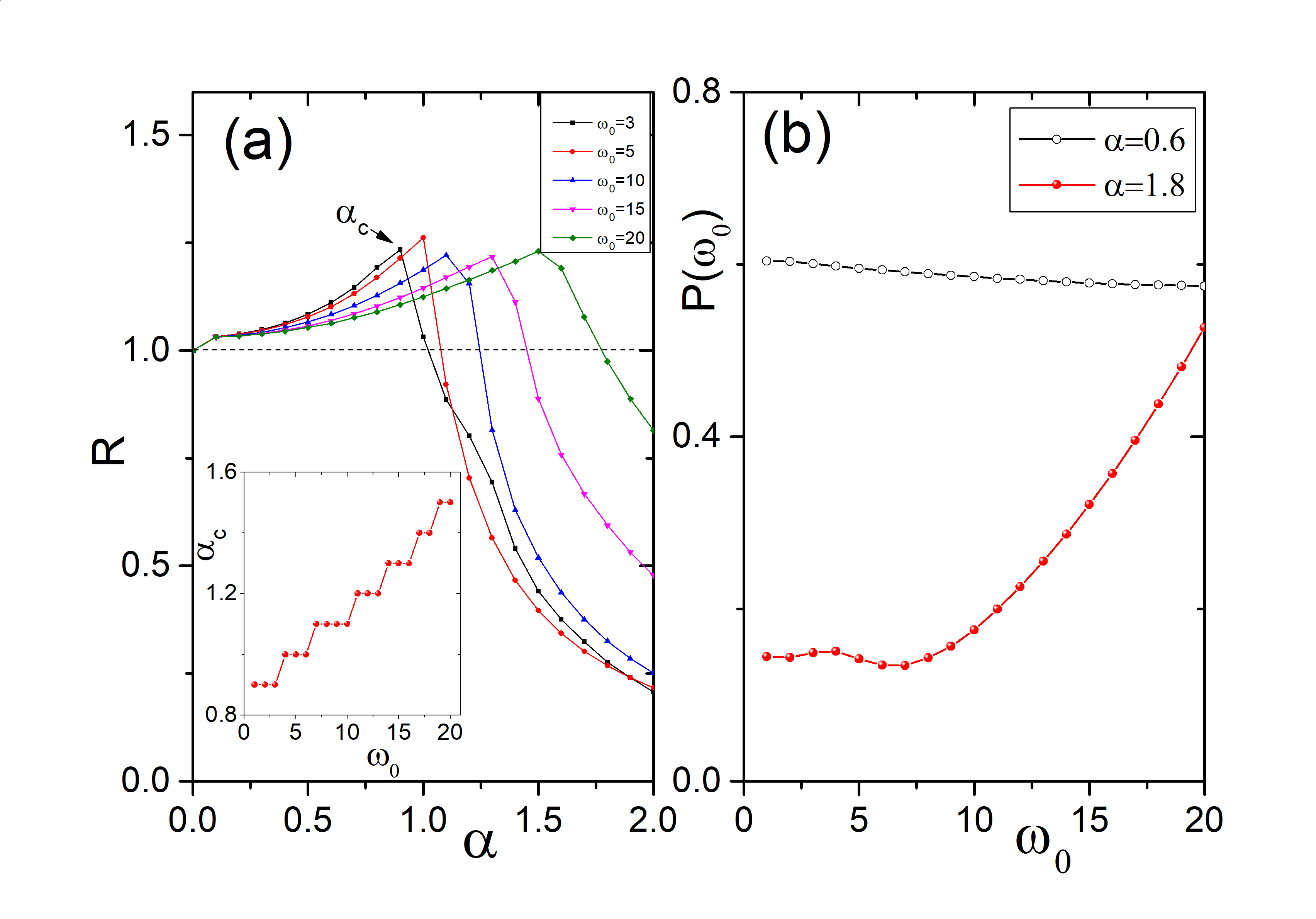}
\caption{(color online) (a) The ratio $R$ defined in the context
versus $\alpha$ for $\omega_0=3,5,10,15,20$, respectively. The inset
presents the critical value $\alpha_c$ versus the modulation
frequency $\omega_0$, where $\alpha_c$ is the value when $R$ gets
the maximum (the coupled oscillators have largest AD domains in
parameter space $(\epsilon_0, \Delta\omega))$ (b) The proportion
$P(\omega_0)$ of AD states versus $\omega_0$ for a given
$\alpha=0.6$ (black hollow-dotted line),$\alpha=1.8$ (red
solid-dotted line), respectively.} \label{fig_5}
\end{figure}

To integrally figure out how the periodically modulated coupling
influences the AD domain of the coupling oscillators, it is natural
to reveal the mechanical regimes of the stabilities of AD under the
periodically modulated coupling. Generally the stability of the AD
domain in coupled oscillators is obtained from linear stability
analysis of Eq. \ref{eq1} around $Z_i = 0$ if the coupling strength
is constant. The characteristic eigenvalue is \cite{aro}
\begin{eqnarray}\label{eq3}
\lambda_{1,2,3,4}&=&1-\epsilon\pm\sqrt{\epsilon^2-\frac{(\omega_2-\omega_1)^2}{4}}\pm
j \frac{(\omega_1+\omega_2)}{2},
\end{eqnarray}
then the AD domain is determined by $Re(\lambda)<0$, that is,
$1\le\epsilon\leq\Delta\omega/2$ and $\Delta\omega/2<\epsilon<1/2+\
\Delta\omega^2/8$ which is right the boundary lines of the AD domain
as shown in Fig. \ref{fig_3}(a). However, when $\epsilon$ is varying
with time, the linear stability analysis around the original fixed
points is not available any more. We numerically record the AD
domain in parameter space $\Delta \omega$ versus $\epsilon_0$ (both
are in the range of [0,20]) for a given $\omega_0=4.0$, and
$\alpha=0, 0.5,1.0, 1.1,1.4,1.8$ as shown in Figs.
\ref{fig_3}(a)-(f), respectively. When the parameters of the coupled
oscillators are in the blue (red) area, the oscillators are in AD
(oscillating) states. With the increment of the modulation amplitude
$\alpha$, the AD domain firstly expands by decreasing the critical
frequency mismatches, which is the lower boundary of the AD domain
for given average coupling strength $\epsilon_0$. Then the AD domain
shrinks with the increment of $\alpha$ when $\alpha$ is larger than
a critical value $\alpha_c$. Moreover, the AD domains split into two
parts when $\alpha>1$ (it is repulsively coupled in some interval of
each period) leading to a kind of ragged AD along the direction of
frequency mismatch $\Delta\omega$. It should be emphasized that the
AD is firstly observed to be ragged in the direction of parameter
space $\Delta\omega$, as a comparison, OD domain is ragged in the
direction of parameter space $\epsilon$ in the coupled system with a
certain spatial frequency distribution \cite{ma}. There are two
segments of AD domains in parameter space of $\Delta\omega$ versus
$\epsilon_0$, as an example, AD state occurs in the intervals of
$\Delta\omega\in[3.6,5.4]$ and $\Delta\omega\in[9.3,20]$ as
$\alpha=1.1$ and $\epsilon_0=5$, which can be also seen from the
vertical line in Fig. \ref{fig_3}(d). The two ragged AD domains keep
shrinking and leaving away from each other with the increment of the
modulation amplitude $\alpha$. Meanwhile, for a given $\alpha=1.5$,
we explore the effects of the modulation frequency $\omega_0$ by
numerically plotting the phase diagram in parameter space $\Delta
\omega$ versus $\epsilon_0$ for $\omega_0=1,5,10,13,16,19$ in Figs.
\ref{fig_4}(a)-(f), respectively. The results show that the
increment of the modulation frequency $\omega_0$ firstly splits the
AD domain into two parts with the upper one larger than the lower
one, then the lower larger one expands while the upper larger one
shrinks. Finally, the two parts merged into one large AD domain
again.
 To present detailed insight of the effects of $\alpha$ on the AD domain, the
normalized ratio factor $R=S(\alpha)/S(\alpha=0)$ is defined to
qualify the change of AD domains under designated regions
($\epsilon_0\in[0,20]$ and $\delta\omega\in[0,20]$) in Fig.
\ref{fig_5}(a) where the modulation frequency
$\omega_0=3,5,10,15,20$, respectively. $S(\alpha)$  is the area of
the AD domains for given $\alpha$ while $S(\alpha=0$) represents the
area of AD domains for $\alpha=0$ (i.e., the red domains in Fig.
\ref{fig_3}(a)).

It is obvious that the ratio $R$ firstly increases slightly
(expansion of AD domain) and then decreases sharply to small value
(reduction of AD domain) as $\alpha$ increases from 0 to 2 for all
modulation frequencies $\omega_0$. There is a critical value
$\alpha_c$ with which the AD domain gets to the largest value for
each given modulation frequency $\omega_0$. That is to say there is
an optimal modulation amplitude of the coupling strength with which
the coupled system has the largest AD domain. When
$\alpha<\alpha_c$, the increment of $\alpha$ tends to enlarge the
area of AD domain by shrinking that of the oscillating domain,
otherwise, the AD domain is torn into multi domains by the birth of
oscillating domain and shrinks with the increment of modulation
amplitude $\alpha$. Interestingly, the optimal value of $\alpha_c$
increases with the increment of the modulation frequency $\omega_0$
as shown the inset in Fig. \ref{fig_4}(a). The larger the modulation
frequency $\omega$ is, the larger the modulation amplitude is needed
to maximize the AD domain.

Now let's focus on the effects of the modulation frequency
$\omega_0$ on the AD domain. Define the proportion of the AD domain
on the designated regions $\epsilon_0\in[0,20]$ and
$\delta\omega\in[0,20]$ as $P(\omega_0)=S(\omega_0)/S_{tot}$ for
given $\omega_0$, where $S(\omega_0)$ is the area of the AD domain
for given $\omega_0$, and $S_{tot}$ is the area of the designated
region. Then $P(\omega_0)$ versus the modulation frequency
$\omega_0$ can be numerically presented for an arbitrarily given
modulation amplitude $\alpha$. Figure. \ref{fig_5}(b) presents the
results of $P(\omega_0)$ versus $\omega_0$ for given
$\alpha=0.6,1.8$ respectively. $P(\omega_0)$ linearly decreases with
a slow speed for $\alpha=0.6$ while obviously increases  for
$\alpha=1.8$ as $\omega_0$ increase from 7 to 20. Therefore, the
modulation frequency $\omega_0$ is beneficial to shrink  AD domain
for small modulation amplitude $\alpha$, otherwise, the AD domain
expands quickly with the increment of the modulation frequency
$\omega_0$ as $\alpha$ is large (larger than the maximal $\alpha_c$.

\subsection{ B. Coupled Rossler oscillators}

AD can also be observed in coupled chaotic oscillators with time
delay coupling \cite{pra}, and on-off coupling schemes \cite{sun}.
It is nature to explore the effects of the periodically modulated
coupling on the AD domain in coupled chaotic nonidentical
oscillators. Let's consider the coupled chaotic Rossler oscillators
with two different time scales.
\begin{eqnarray}\label{eq4}
\dot{x_i}(t)&=&-\omega_i(y_i(t)+z_i(t)),\nonumber \\
\dot{y_i}(t)&=&\omega_i(x_i(t)+a y_i(t)),\nonumber \\
\dot{z_i}(t)&=&\omega_i(b+z_i(t)(x_i(t)-c).
\end{eqnarray}

\begin{figure}
\includegraphics[width=9cm]{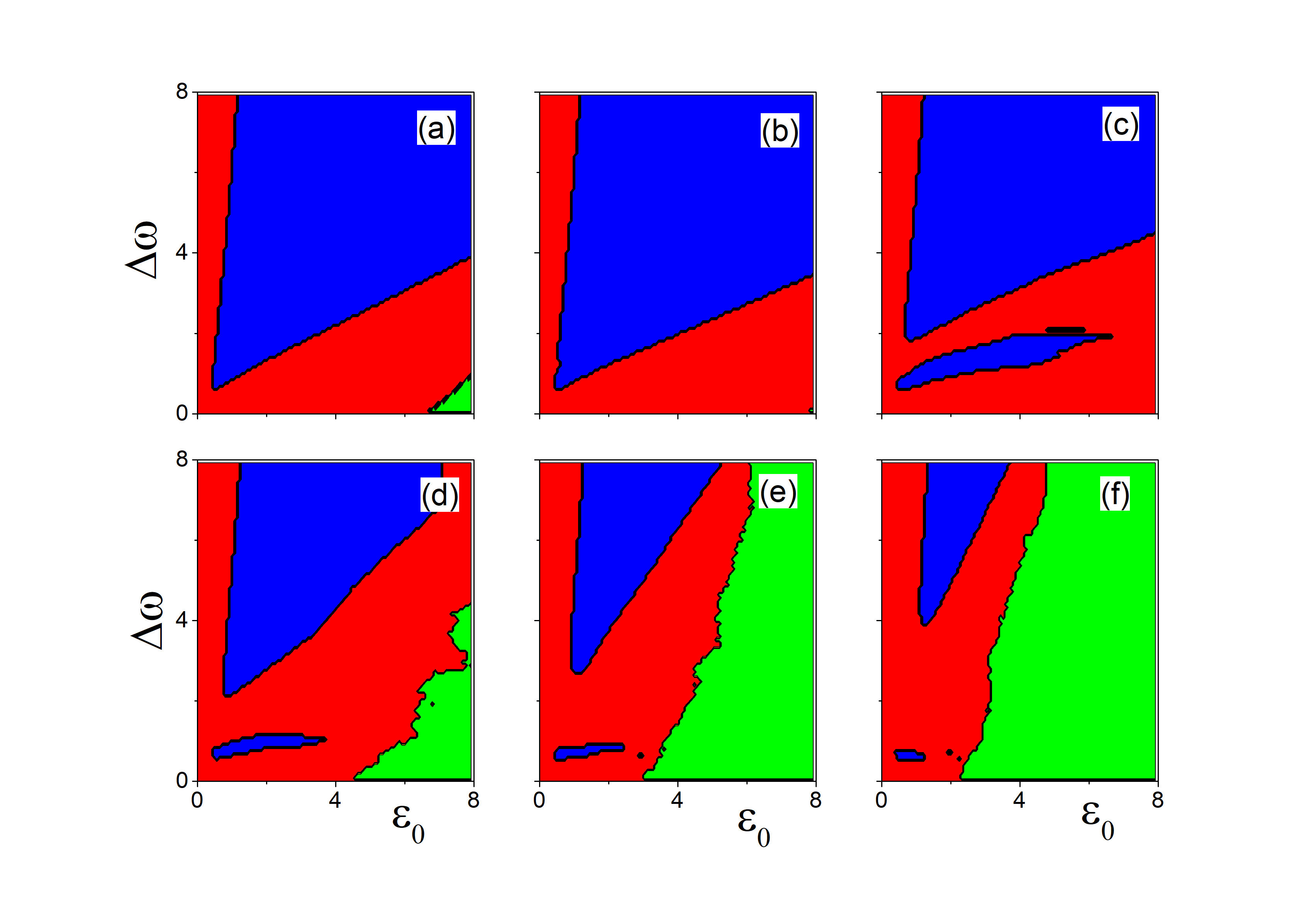}
\caption{ (color online) The phase diagram of the parameters
$\Delta\omega$ versus $\epsilon_0$  in coupled Rossler oscillators
for (a) $\alpha=0.0$, (b) $\alpha=0.5$, (c) $\alpha=1.0$, (d)
$\alpha=1.2$, (e) $\alpha=1.4$, (f) $\alpha=1.6$, respectively,
where the blue domain is AD state, the red domain is the oscillating
state, and the blue domain is the blowup state.} \label{fig_6}
\end{figure}

where $\omega_i$ rescales the rolling speed of single chaotic
oscillator. The single uncoupled oscillator is in chaotic regime for
given parameters $a = 0.15$, $b = 0.4$, $c = 8.5$, and has an
unstable fixed point $(-ay^*,-z^*,z^*)$ with
$z^*=(c-\sqrt{c^2-4ab})/(2a)$. $\omega_1$ is arbitrarily set to 2 ,
and the frequency mismatch of two coupled oscillators is
$\Delta\omega=|\omega_1-\omega_2|$. The periodically modulated
coupling term $\epsilon(t)$ is the same as Eq. \ref{eq2} with the
modulation frequency $\omega_0=1$. The coupling scheme is set as
$\Gamma=\left[
\begin{matrix}
1&0&0\\
0&0&0\\
0&0&0
\end{matrix}\right]$, (i.e. the interacting variable is $x(t)$).
Then the AD domain with different periodic coupling strength can be
conveniently observed by presenting the phase diagram of parameters
$\Delta\omega$ versus $\epsilon_0$ for
$\alpha=0,0.5,1.0,1.2,1.4,1.6$ as shown in Figs. \ref{fig_6}(a)-(f),
respectively. With constant coupling strength ($\alpha=0$), the
coupled Rossler oscillator has three states, AD state (v-shaped blue
domain), oscillating state (red domain), and blowup to infinite
(green domain).  As $\alpha=0.5$ the AD domain is enlarged while the
domain of blowup state is  shrunk. Then the AD domain is ragged into
two parts as $\alpha$ is larger than $1.0$. Finally, the AD domain
shrinks while the domain of the blowup state occurs again and keeps
enlarging as $\alpha$ increases from $1.2$ to $2.0$.

\begin{figure}
\includegraphics[width=9cm]{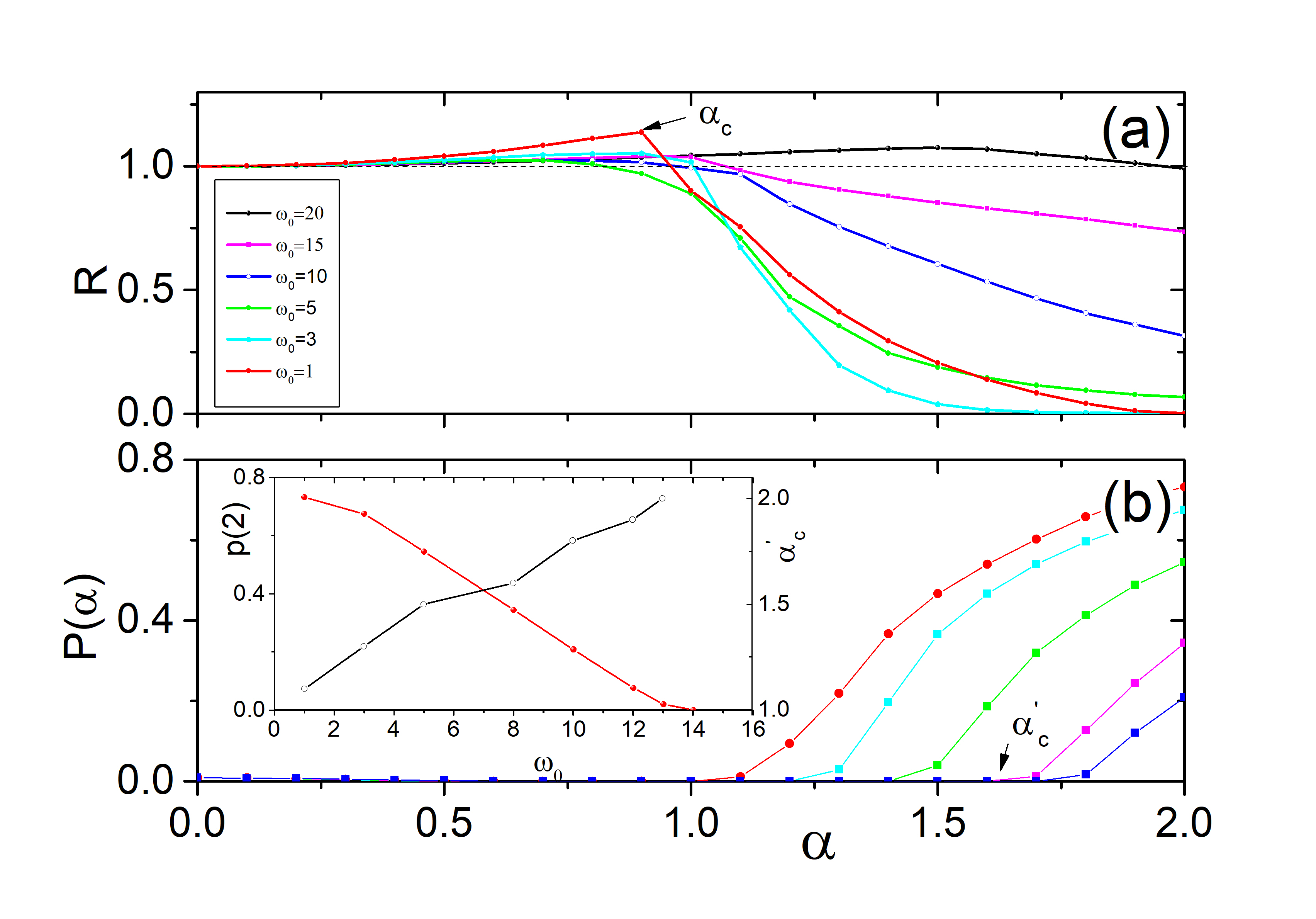}
\caption{(color online) (a) The ratio $R$ defined in the context
versus $\alpha$ for $\omega_0=1,3,5,10,15,20$, respectively. (b) The
proportion $P(\alpha)$ of blowup domains versus $\alpha$ for given
$\omega_0=1$ (red line), $\omega_0=3$ (cyan line), $\omega_0=5$
(green line), $\omega_0=8$ (magenta line), $\omega_0=10$ (blue
line), respectively. The red solid dotted line in inset is the
proportion of  $P(\alpha=2)$ versus $\omega_0$, and the black hollow
dotted line in inset is the critical value of $\alpha'_c$, where
$\alpha'_c$ is the critical value when $p(\alpha)>0$.} \label{fig_7}
\end{figure}

Figure. \ref{fig_7}(a) exhibits the change of areas of AD domain by
the ratio factor $R$ versus the modulation amplitude $\alpha$ for
given modulation frequency $\omega_0=1,3,5,10,15,20$, respectively.
Where the ratio factor $R$ is defined the same as the above one with
the designated region of $(\epsilon,\Delta\omega)$ space in the
range of $\epsilon\in[0,8]$ and $\Delta\omega\in[0,8]$. The effects
of the modulation frequency and modulation amplitude of the coupling
strength on AD domains in coupled Rossler oscillators are similar to
that in the coupled Stuart-Landau oscillators. The increment of the
modulation amplitude tends to enlarge the AD domain first then
shrinks the AD domain for a given modulation frequency $\omega_0$.
There is also a critical value $\alpha_c$ with which the coupled
Rossler oscillators have the largest AD domain. Similarly, we may
defined the proportion of blowup domain as
$P(\alpha)=S(\alpha)/S_{tot}$ for given $\alpha$, where $S(\alpha)$
is the area of parameter space of blowup domain and $S_{tot}$ is the
area of domain in the designated area of $\epsilon\in[0,8]$ and
$\Delta\omega\in[0,8]$. Then the effects of $\alpha$ on blowup can
be indicated by $P(\alpha)$ versus $\alpha$ as shown in Fig.
\ref{fig_7}(b) for given $\omega_0=1,3,5,8,10$, respectively. It is
obvious that $P(\alpha)$ is small and approaches to zero for small
$\alpha$ and then grow up again when $\alpha$ is larger than a
critical value $\alpha'_c$. The critical value $\alpha'_c$ increases
linearly with the increment of the modulation frequency $\omega_0$
as shown the inset in Fig. \ref{fig_7}(b). Meanwhile, the increment
of $\omega_0$ tends to decrease linearly the area of blowup domain
according to the relationship between $p(2)$ versus $\omega_0$ for
given $\alpha=2$ as shown the red solid-dotted line in the inset of
Fig.\ref{fig_7}(b).

\section{IV. MECHANISM ANALYSIS}

\begin{figure}
\includegraphics[width=9cm]{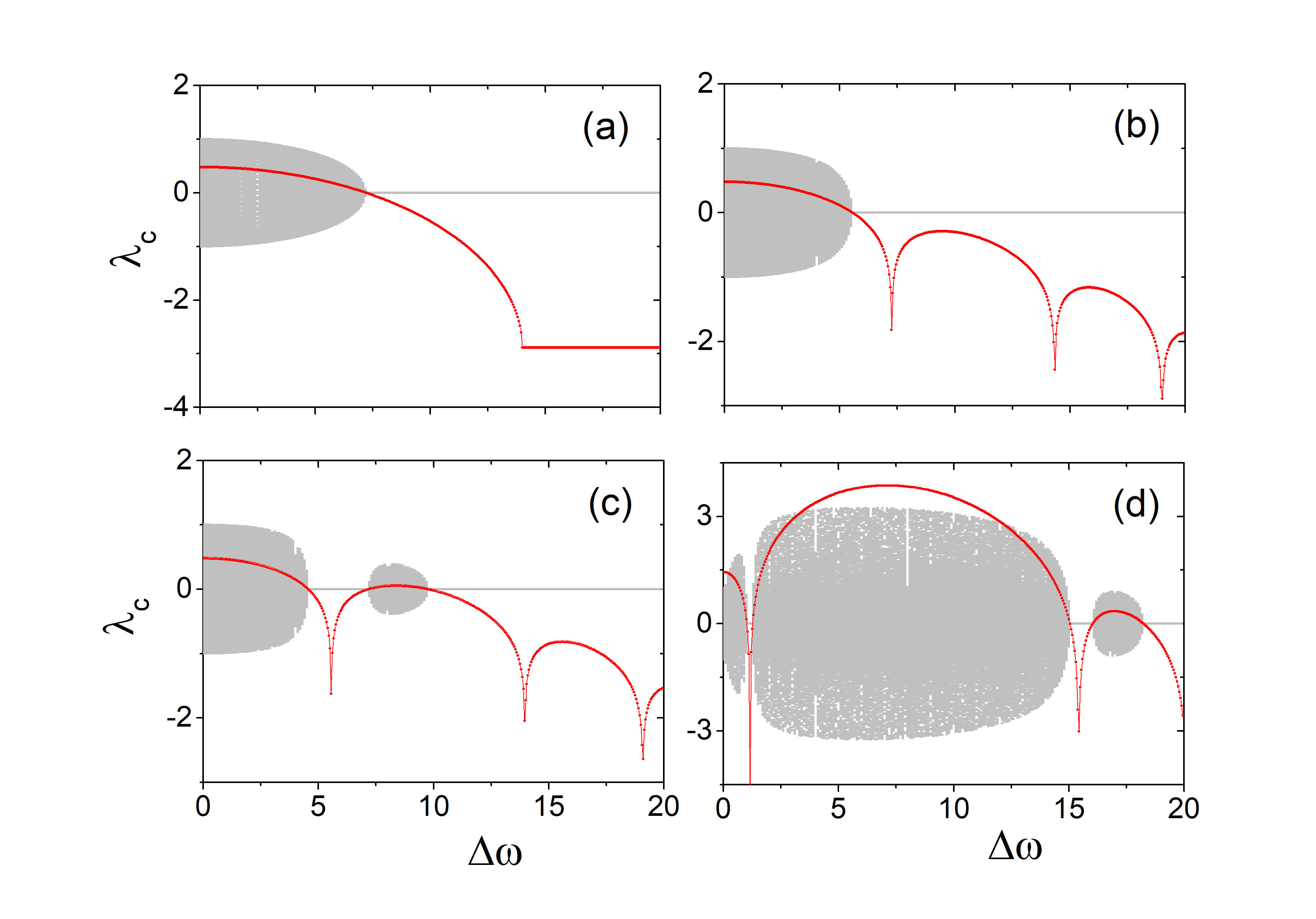}
\caption{(color online) The conditional Lyapunov exponent
$\lambda_c$ versus $\Delta\omega$ for $\epsilon=7$, $\omega_0=4$£¬
and (a) $\alpha=0.0$, (b) $\alpha=0.8$, (c)$\alpha=1.0$, (d)
$\alpha=1.8$, respectively. The grey dots are bifurcation diagram. }
\label{fig_8}
\end{figure}

Since the modulated coupling strength varies with time, the linear
stability analysis near the fixed points is not available to predict
the dynamics regimes. Based on the fact that the conditional
Lyapunov exponent \cite{pyr} is a valid tool to determine the
generalized synchronization, it is expected to determine the
stability of the AD state in the coupled nonidentical oscillators.
Let $\delta z_i=z_i-z^*_i,(i=1,2)$ be an infinitesimal perturbation
added to oscillator $i$, then whether the perturbed trajectories of
Eq. \ref{eq1} could be converged to the fixed point $z^*$ is mainly
determined by the set of variational equations
\begin{eqnarray}\label{eq5}
\left[ \begin{matrix}
 \dot{\delta z_1}(t)  \\
   \dot{\delta z_2}(t)
  \end{matrix}\right]=
\left[ \begin{matrix}
 (DF_1(z^*_1))&0  \\
   0&DF_2(z^*_2) \end{matrix}\right]
\left[ \begin{matrix}
\delta z_1(t)  \\
\delta z_2(t)  \\
\end{matrix}\right]+\epsilon(t) \Gamma A
\left[ \begin{matrix}
\delta z_1(t)  \\
\delta z_2(t)  \\
\end{matrix}\right]
\end{eqnarray}
Here $z^*_1=(0,0)$, $z^*_2=(0,0)$ are the original fixed points of
single oscillators. $DF_1()$ and $DF_2()$ are the deviations of the
two coupled oscillators.  $\Gamma$ is the the coupling scheme
($\Gamma=\left[ \begin{matrix}
1&0\\
0&1
\end{matrix}\right]$ for coupled LS oscillators) and
$A=\left[ \begin{matrix}
-1&1\\
1&-1
\end{matrix}\right]$ is the link matrix whose eigenvalue is $\lambda_1=0$
and $\lambda_2=-2$. Solving Eq. \ref{eq5} numerically for
$\lambda_2=-2$, we are able to obtain the conditional Lyapunov
exponent $\lambda_c$ with respect to the parameters  of the coupling
strength and frequency mismatches, based on which the stable domain
of AD, i.e. the domain with $\lambda_c<0$, can be identified. In
Figs. \ref{fig_8}(a)-(d), we plot the conditional Lyapunov exponent
$\lambda_c$ as a function of $\Delta\omega$ for $\epsilon_0=7$, and
$\alpha=0.0,0.8,1.0,1.8$, respectively. The conditional Lyapunov
exponent transit to negative from positive when coupled system
transit from oscillating state to AD state which matches well with
the bifurcation results in grey dots.

\begin{figure}
\includegraphics[width=9cm]{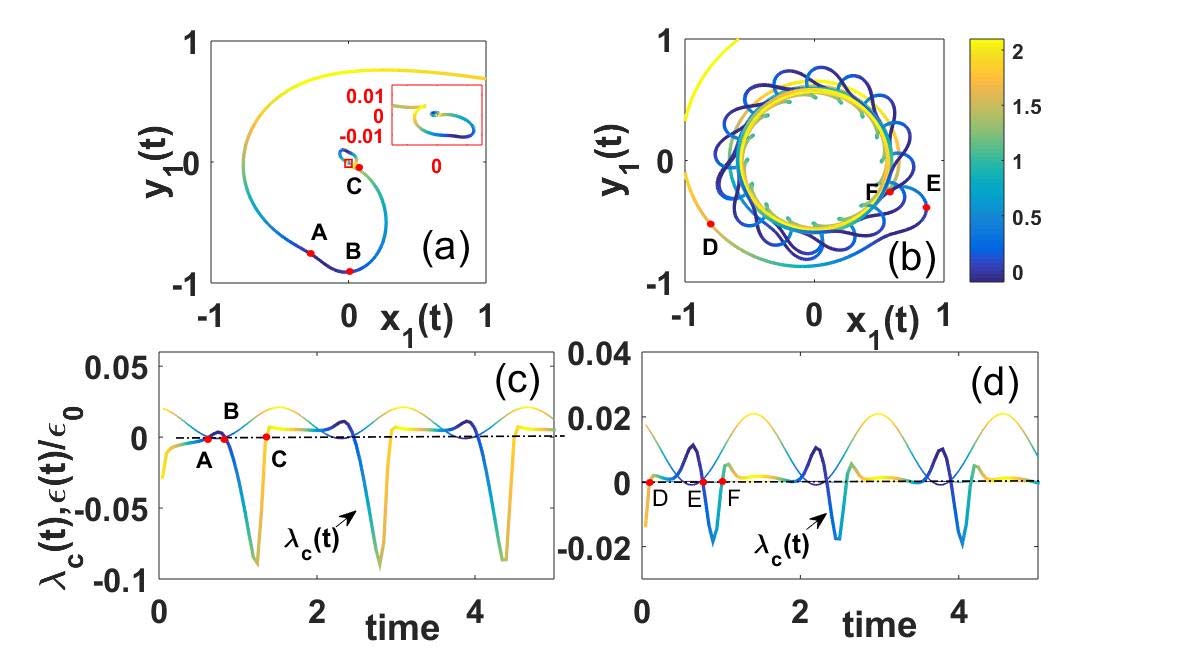}
\caption{(color online) (a) The phase diagram of $y_1(t)$ versus
$x_1(t)$ for $\alpha=1.1$, $\epsilon=5$, $\Delta\omega=4.5$. The
inset is the enlarged part of the red squared area. A,B,C note the
position when the local conditional Lyapunov exponent $\lambda_c(t)$
is crossing $x$ axis as shown in Fig. \ref{fig_9}(c). The colors of
lines in the Fig. \ref{fig_9} denote the values of the normalized
coupling strength $\epsilon(t)/\epsilon_0$. (b) The phase diagram of
$y_1(t)$ versus $x_1(t)$ for $\alpha=1.1$, $\epsilon=5$,
$\Delta\omega=6.5$. D,E,F note the position when the local
conditional Lyapunov exponent $\lambda_c(t)$ is crossing $x$ axis as
shown in Fig. \ref{fig_9}(d). The local conditional Lyapunov
exponent and the normalized coupling strength ($1\%$ of
$\epsilon(t)/\epsilon_0$) versus time for (c) $\Delta\omega=4.5$ and
(d) $\Delta\omega=6.5$} \label{fig_9}
\end{figure}

\begin{figure}
\includegraphics[width=9cm]{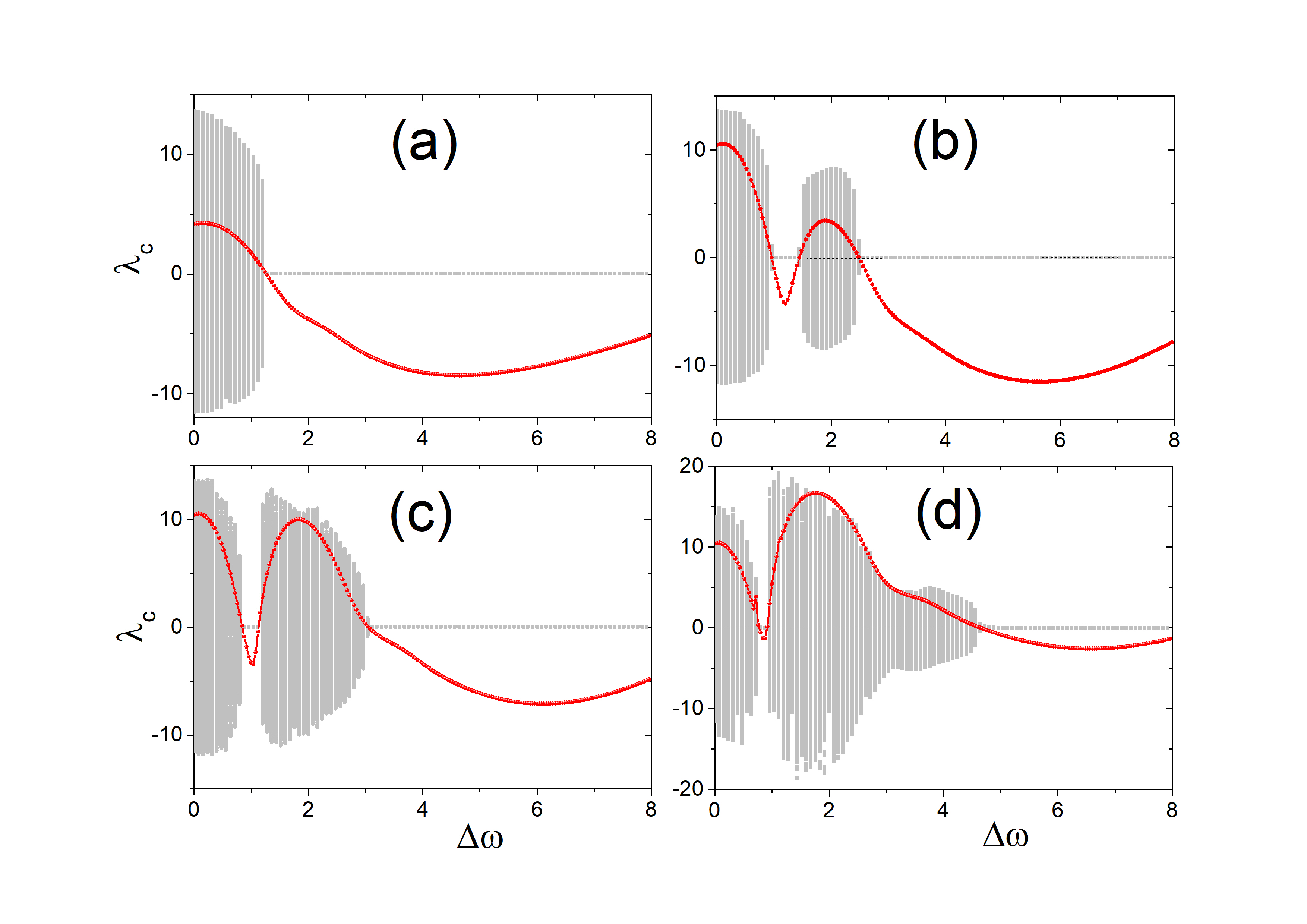}
\caption{(color online)  The conditional Lyapunov exponent
$\lambda_c$ (red lines) versus $\Delta\omega$ for  $\epsilon=2$ and
(a) $\alpha=0.5$, (b) $\alpha=1.0$, (c)$\alpha=1.2$, (d)
$\alpha=1.4$, respectively in the coupled Rossler oscillators (the
conditional Lyapunov exponent is magnified 50 times for the
convenience of observation in the figure). The grey dots are
bifurcation diagram. } \label{fig_10}
\end{figure}

To observe clearly  how the varying coupling strength influences the
dynamical of the coupled oscillators, the phase diagrams of $x_1(t)$
versus $y_1(t)$ for $\Delta\omega=4.5$ (AD state in numerical
results), and $\Delta\omega=6.5$ (oscillating state in numerical
results) are presented  in Figs. \ref{fig_9}(a)(b), respectively.
The value of the normalized coupling strength
($\epsilon(t)/\epsilon_0$) is indicated by the color of the phase
diagram. Fig. \ref{fig_9}(a) indicates that the oscillator may leave
away from (stage AB)  or approach to (stage BC) the original fixed
point ($0,0$) in each period of the modulated coupling strength. The
speed of approaching to the original fixed point is larger than that
of leaving away which results to that the coupled system approaches
to the fixed points gradually as time goes to infinite. The enlarged
diagram indicates that the coupled system may leave away from the
original point in some interval of each period of the modulated
coupling  after approaching to the fixed point. Therefore, the
oscillator will never stay on the original fixed point no matter how
near it approach to the original fixed point (in this sense, the
original fixed point is not stable). However, owing to the finite
preciseness of the computer, AD can be observed in the numerical
results when the distance between the oscillator and the original
fixed point is smaller than the computer's preciseness. In Fig.
\ref{fig_9}(b), the oscillator may also diverge from (stage DE) and
approach to (stage EF) the original fixed point in each period of
the  modulated coupling. Noted that the speed of approaching to
fixed point is larger than that of diverging from the fixed point,
however, the time of the former one is shorter than the later one.
As a result, the coupled oscillator forms an
 oscillating state. The speed of approaching to or leaving away from
 the fixed point is related to the value of the local conditional Lyapunov exponent \cite{pik1}  as shown in Figs.
\ref{fig_9}(c)(d), respectively. Positive local conditional Lyapunov
exponent makes the oscillator leave away from the fixed point while
negative one drive the coupled system to converge to the fixed
point. The speed of approaching to or diverging from the fixed point
is related to the absolute value of the local conditional Lyapunov
exponent. Compared the results between Fig. \ref{fig_9}(c) and Fig.
\ref{fig_9}(d), the final fate of the coupled oscillator is
completely determined by the average value of the local conditional
Lyapunov exponent in one modulation period. The coupled system is in
AD state if the average value of the local conditional Lyapunov
exponent is negative, otherwise it is in oscillating state. The
conditional Lyapunov exponent is also available to predict the AD
dynamics in the coupled Rossler oscillators. Figs.
\ref{fig_10}(a)-(d) present the conditional Lyapunov exponent
together with the bifurcation diagram on variable $x_1$ versus
$\Delta\omega$. The conditional Lyapunov exponent  gets to negative
value when the coupled Rossler oscillators experience AD which also
agrees well with the bifurcation diagram.

To verify the stability of the fixed points further, we add the
noise on the coupled system,

\begin{eqnarray}\label{eq6}
\dot{X_1}(t)&=&f_1(X_1(t))+\epsilon(t)\Gamma(X_{2}(t)-X_{1}(t))+\xi_1(t),
\nonumber \\
\dot{X_2}(t)&=&f_2(X_2(t))+\epsilon(t)\Gamma(X_{1}(t)-X_{2}(t))+\xi_2(t),
\end{eqnarray}
 where the independent stochastic variables
$\xi_i(t)$ are the zero-mean white Gauss noises with strength
$\sigma$, namely,
$<\xi_i(t)>=0$,$<\xi_i(t)\xi_j(t')>=2\sigma\delta(t-t')\delta_{ij}$,$\gamma$
is the same as the above. If the stability of AD is strong, then the
periodically modulated coupled system with noise may meander in the
vicinity of fixed point. Define the variable $\gamma$ as the ability
of resisting noise as following,
\begin{equation}\label{eq7}
\gamma=\left\{\begin{array}{lr}
50,& \eta>=20,\\
40,& 15\leq\eta<20 \\
30,& 10\leq \eta<15 \\
20,& 2\leq\eta<10\\
10,&  \eta<2,
\end{array}
\right.
\end{equation}
where $\eta=max(x_1-x^*_1)/\sigma$ is the ratio of the maximum
$x_1(t)-x^*_1$ to the noise strength $\sigma$. Then AD is more
stable if $\eta$ is smaller. The phase diagram of $\gamma$ in the
parameter $\Delta\omega$ versus $\epsilon_0$ of coupled LS
oscillators are presented in Figs. \ref{fig_10} (a)-(b) for
$\alpha=1.1$, and $\omega_0=4$, as well as that of coupled Rossler
oscillators presented in Figs. \ref{fig_10} (c)-(d) for
$\alpha=1.0$, and $\omega_0=1$. According to Figs.
\ref{fig_10}(a)(b), it is easy to conclude that the ragged AD domain
in the upper part (except the edge parts) has stronger stability
than that in the lower part. Noticed that the upper part of the
ragged AD domain is stable for constant coupling strength (without
modulation) while the lower AD domain is newly born by the modulated
coupling which has less stability than the original AD domain.
Moreover, the conclusion is slightly influenced by the noise
intensity.

\begin{figure}
\includegraphics[width=9cm]{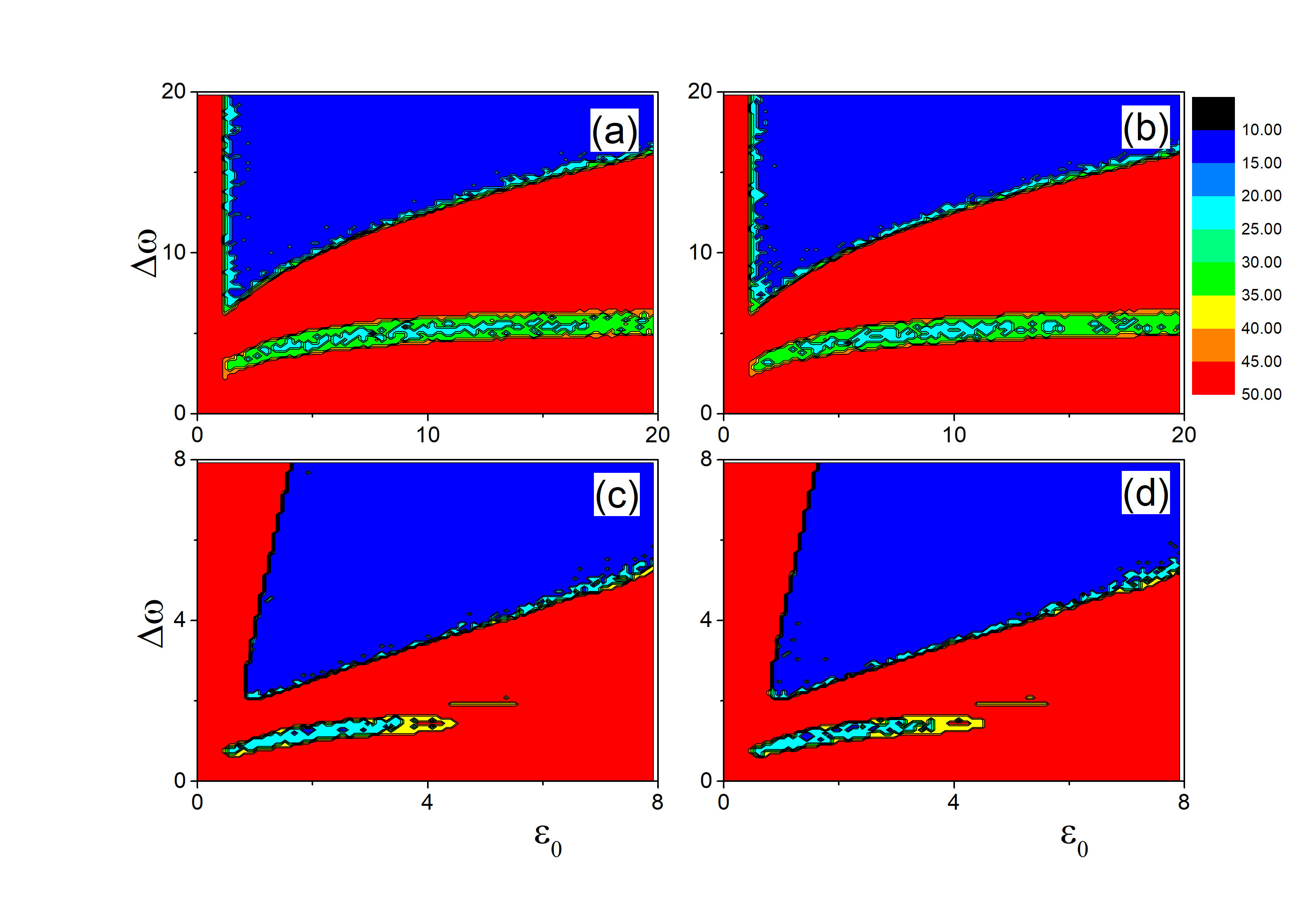}
\caption{The phase diagram of the parameters $\Delta\omega$ versus
$\epsilon_0$ with noise strength being (a) $\sigma=0.001$ in coupled
$SL$ oscillators, (b) $\sigma=0.1$ in coupled SL oscillator,(c)
$\sigma=0.001$ in coupled Rossler oscillators, (d) $\sigma=0.1$ in
coupled Rossler oscillators. The colorbar determines the values of
$\gamma$ in Eq. \ref{eq7}.} \label{fig_11}
\end{figure}

\textbf{V. Conclusions}

Totally, both the modulation frequency and the modulation amplitude
of the periodically modulated coupling strength significantly
influenced the dynamics of the coupled limit cycles or chaotic
oscillators. The increment of the modulation amplitude firstly
increases the AD domain and then decreases the AD domain as it is
larger than a critical value which is related to the modulation
frequency. That is to say, small modulation amplitude of the
coupling strength is helpful to enlarge the AD domain of the coupled
nonidentical oscillators. However, when the coupling term is varying
between repulsive (negative) and attractive (positive), the AD
domains may be shrunk and ragged to several parts by the occurrence
of oscillating domain. Meanwhile, the increment of modulation
frequency of the periodic coupling tends to slightly decrease the AD
domains for small modulation amplitude while dramatically increase
the AD domains for large modulation amplitude. According to the
local conditional Lyapunov exponent of the periodically coupled
oscillators, one may find that the stability of the AD states is
varying with the coupling strength. Whether the coupled oscillators
can converge to AD state or not is completely determined by the sign
of the averaged conditional Lyapunov exponent. The periodically
modulated coupling is beneficial to realize AD and is more easy to
physical realization than on-off coupling which needs high speed
stitchers and is difficult to apply, therefore, it has potential
application in the dynamical control in engineering.

\textbf{Acknowledgement} Weiqing Liu is supported by the National
Natural Science Foundation of China (Grant No. 11765008) and the
Qingjiang Program of Jiangxi University of Science and Technology.
Jiangnan Chen is supported by the Project of Jiangxi province.


\end{document}